\documentclass{aa}  

\usepackage{graphicx}
\usepackage{txfonts}
\usepackage{hyperref}  
\hypersetup{
  colorlinks=true,  
  urlcolor=blue,    
  linkcolor=red,
  citecolor=blue     
}

\usepackage{natbib}
\bibpunct{(}{)}{;}{a}{}{,} 

\usepackage{threeparttable}

\defcitealias{1990gan}{GF90}
\defcitealias{2012carlsson}{CL12}

\begin{document}

   \title{An approximate recipe of chromospheric radiative losses for solar flares}

   \author{Jie Hong
          \inst{1,2}
          \and
          Mats Carlsson\inst{3,4}
          \and
          M. D. Ding\inst{1,2}
          }

   \institute{School of Astronomy and Space Science, Nanjing University, Nanjing 210023, PR China\\
              \email{jiehong@nju.edu.cn}
         \and
             Key Laboratory for Modern Astronomy and Astrophysics (Nanjing University), Ministry of Education, Nanjing 210023, PR China
          \and
          Rosseland Centre for Solar Physics, University of Oslo, P.O. Box 1029 Blindern, NO-0315 Oslo, Norway
          \and
           Institute of Theoretical Astrophysics, University of Oslo, P.O. Box 1029 Blindern, NO-0315 Oslo, Norway
             }

   \date{}

 
  \abstract
   {Radiative losses in the chromosphere are very important in the energy balance. There have been efforts to make simple lookup tables for chromospheric radiative losses in the quiet Sun. During solar flares, the atmospheric conditions are quite different, and the currently available recipe of \citet{1990gan} is constructed from semi-empirical models. It remains to be evaluated how these recipes work in flare conditions.}
   {We aim to construct an approximate recipe of chromospheric radiative losses for solar flares.}
   {We follow the method of \citet{2012carlsson} to tabulate the optically thin radiative loss, escape probability, and ionization fraction,  while using a grid of flare models from radiative hydrodynamic simulations as our dataset.}
   {We provide new lookup tables to calculate chromospheric radiative losses for flares. Compared with previous recipes, our recipe provides a better approximation to the detailed radiative losses for flares.}
   {}

   \keywords{Radiative transfer  --
                 Sun: chromosphere --
                 Sun: flares
               }

   \maketitle
%

\section{Introduction}
In the solar atmosphere, radiation plays an important role in the energy balance. The contribution of radiation to the change in the internal energy, often referred to as radiative losses, is quantified as the divergence of the radiative flux:
\begin{equation}
\label{ }
Q_\mathrm{rad}=-\nabla\cdot\mathcal{F}.
\end{equation}
A positive value of $Q_\mathrm{rad}$ means that there is local heating from the extinction of photons, and a negative value indicates local cooling through the emission of photons.

In the transition region and corona, the atmosphere is optically thin and the radiative losses can be simplified in the form of
\begin{equation}
\label{ }
Q_\mathrm{rad}=-\Lambda(T,n_\mathrm{e})n_\mathrm{e}n_\mathrm{H},
\end{equation}
where $n_\mathrm{e}$ is the electron density, $n_\mathrm{H}$ is the hydrogen density, and $\Lambda$ can be calculated under the coronal approximation with the CHIANTI database \citep{1997dere,2021delzanna}.
   
However, in the chromosphere, the strong lines are normally optically thick, which means that there is a probability for the energy (in the form of photons) to escape. \citet[hereafter \citetalias{2012carlsson}]{2012carlsson} wrote the radiative loss function from element $X$ at ionization stage $m$ as:
\begin{equation}
\label{cl12}
Q_{\mathrm{rad},X_m}=-L_{X_m}E_{X_m}\frac{n_{X_m}}{n_X}\frac{n_X}{n_\mathrm{H}}\frac{n_\mathrm{H}}{\rho}n_\mathrm{e}\rho,
\end{equation}
where $L_{X_m}$ is the thin radiative loss function, $E_{X_m}$ is the escape probability and $\frac{n_{X_m}}{n_X}$ is the ionization fraction of element $X$ at ionization stage $m$. These three parameters are determined empirically from radiative (magneto)hydrodynamic simulations of the quiet Sun. $\frac{n_X}{n_H}$ is the abundance of element $X$ relative to hydrogen, $\frac{n_\mathrm{H}}{\rho}$ is the number of hydrogen particles per mass unit, a constant dependent on abundances, and $\rho$ is the mass density. This approximate recipe can reproduce radiative cooling of the quiet Sun very well, and has been included in many radiative (magneto)hydrodynamic codes \citep[e.g.][]{2011gudiksen,2013bradshaw,2020wang}.

During solar flares, the chromosphere undergoes drastic changes, with a rapid rise in temperature, electron density and pressure. The bombardment of the non-thermal electrons can also increase the excitation and ionization rate of neutral hydrogen in the ground level \citep{1993fang}. With such different local conditions, it is noted that the chromospheric radiative losses in flares are much larger than those in the quiet Sun \citep{1980machado,1986avrett}. The work of \citet[hereafter \citetalias{1990gan}]{1990gan} provides a revised recipe of \citet{1980nagai} based on fitting the radiative loss curves of semi-empirical models. The recipe of \citetalias{1990gan} has a similar form:
\begin{equation}
\label{ }
Q_\mathrm{rad}=-f(T)\alpha(z)n_\mathrm{H}n_\mathrm{e},
\end{equation}
where $f(T)$ is the thin radiative loss function, and $\alpha(z)$ is the probability that the energy escapes from height $z$.

\begin{table*}
\caption{Line fluxes and total radiative fluxes of the chromosphere from lines and Lyman continuum in different flare models. }             
\label{flux}      
\centering
\resizebox{\textwidth}{!}{
\begin{tabular}{c | rrccr  | crrc | rrr | c}
\hline
flare &    \multicolumn{5}{c|}{$\mathcal{F}_\mathrm{avg}$ ($10^7$ erg cm$^{-2}$ s$^{-1}$)}  & \multicolumn{4}{c|}{$\mathcal{F}^\mathrm{tot}$ ($10^9$ erg cm$^{-2}$)} & \multicolumn{3}{c|}{$\Delta \mathcal{F}^\mathrm{tot}$ (\%)} &flare \\ 
model &    Ly$\alpha$ & H$\alpha$ & \ion{Ca}{ii} K & \ion{Ca}{ii} 8542 \r{A} & \ion{Mg}{ii} k  & detailed & \citetalias{1990gan} & \citetalias{2012carlsson} & this work & \citetalias{1990gan} & \citetalias{2012carlsson} & this work &class$^1$ \\ \hline
\multicolumn{14}{c}{Models for fitting}\\ \hline
f10E05d3 &        5.51 & 3.42 & 0.87 & 0.45 & 2.55 & 5.84  & 7.60  & 57.29  & 4.84  & 30.30  & 881.55  & -17.01 &C2.7 \\ 
f10E10d3 &        7.07 & 2.92 & 0.61 & 0.35 & 2.28 & 5.48  & 13.67  & 78.15  & 5.88  & 149.37  & 1325.90  & 7.34  &...\\ 
f10E15d3 &        6.39 & 2.85 & 0.65 & 0.38 & 1.71 & 5.48  & 9.34  & 86.55  & 6.80  & 70.53  & 1480.13  & 24.06  &...\\ 
f10E20d3 &        4.66 & 2.99 & 0.68 & 0.40 & 1.79 & 5.79  & 7.98  & 100.48  & 6.71  & 37.95  & 1636.76  & 15.93  &...\\ 
f10E25d3 &        3.44 & 3.08 & 0.71 & 0.42 & 1.86 & 5.19  & 7.50  & 89.74  & 5.13  & 44.72  & 1630.54  & -1.06  &...\\ 
f10E05d4 &        3.69 & 3.22 & 0.88 & 0.44 & 2.28 & 4.87  & 6.59  & 37.22  & 3.96  & 35.25  & 663.58  & -18.67  &C7.6\\ 
f10E10d4 &        6.96 & 3.02 & 0.73 & 0.39 & 2.56 & 6.15  & 12.22  & 76.78  & 6.56  & 98.93  & 1149.53  & 6.68  &A8.9\\ 
f10E15d4 &        7.67 & 2.49 & 0.54 & 0.34 & 1.71 & 5.19  & 9.70  & 81.54  & 7.64  & 86.92  & 1471.62  & 47.35  &...\\ 
f10E20d4 &        6.46 & 2.69 & 0.57 & 0.35 & 1.77 & 6.36  & 8.21  & 108.33  & 9.10  & 28.99  & 1602.69  & 43.00  &...\\ 
f10E25d4 &        4.71 & 2.83 & 0.58 & 0.36 & 1.78 & 6.26  & 7.36  & 106.08  & 6.38  & 17.66  & 1595.74  & 1.91  &...\\ 
f10E05d5 &        3.23 & 2.90 & 0.81 & 0.41 & 2.02 & 4.12  & 6.08  & 26.29  & 3.24  & 47.38  & 537.83  & -21.48  &C9.2\\ 
f10E10d5 &        6.45 & 3.05 & 0.77 & 0.40 & 2.56 & 6.14  & 11.55  & 71.84  & 6.46  & 88.16  & 1070.65  & 5.26  &B4.5\\ 
f10E15d5 &        8.10 & 2.25 & 0.49 & 0.32 & 1.74 & 4.70  & 9.77  & 73.06  & 7.49  & 107.63  & 1453.28  & 59.19  &...\\ 
f10E20d5 &        6.90 & 2.49 & 0.52 & 0.33 & 1.69 & 6.23  & 8.33  & 83.67  & 9.20  & 33.70  & 1242.40  & 47.67  &...\\ 
f10E25d5 &        5.49 & 2.64 & 0.52 & 0.33 & 1.70 & 6.70  & 7.16  & 102.43  & 4.20  & 6.92  & 1428.60  & -37.30  &...\\ 
f10E05d6 &        2.83 & 2.66 & 0.76 & 0.39 & 1.83 & 3.61  & 5.72  & 20.54  & 2.74  & 58.61  & 469.43  & -23.89  &C9.8\\ 
f10E10d6 &        6.08 & 3.05 & 0.77 & 0.40 & 2.56 & 5.96  & 11.15  & 66.55  & 6.22  & 87.02  & 1016.10  & 4.35  &C1.4\\ 
f10E15d6 &        8.47 & 2.08 & 0.44 & 0.29 & 1.73 & 4.38  & 10.60  & 62.16  & 7.24  & 141.69  & 1317.65  & 65.14  &...\\ 
f10E20d6 &        7.45 & 2.36 & 0.50 & 0.32 & 1.72 & 6.07  & 8.40  & 71.96  & 7.76  & 38.40  & 1086.18  & 27.93  &...\\ 
f10E25d6 &        6.06 & 2.53 & 0.50 & 0.31 & 1.62 & 6.87  & 7.05  & 96.32  & 2.94  & 2.62  & 1302.35  & -57.26  &...\\ 
f10E05d7 &        2.56 & 2.52 & 0.73 & 0.38 & 1.70 & 3.29  & 5.52  & 17.28  & 2.41  & 67.87  & 425.43  & -26.70  &C9.9\\ 
f10E10d7 &        5.74 & 3.04 & 0.77 & 0.39 & 2.51 & 5.73  & 10.89  & 62.02  & 5.97  & 90.26  & 983.12  & 4.33  &C2.1\\ 
f10E15d7 &        8.51 & 2.04 & 0.39 & 0.28 & 1.74 & 4.07  & 24.17  & 49.62  & 6.99  & 494.22  & 1119.93  & 71.87  &...\\ 
f10E20d7 &        7.97 & 2.28 & 0.48 & 0.31 & 1.70 & 5.85  & 8.23  & 63.71  & 5.05  & 40.67  & 989.04  & -13.69  &...\\ 
f10E25d7 &        6.54 & 2.47 & 0.49 & 0.31 & 1.52 & 6.89  & 6.99  & 89.27  & 2.35  & 1.48  & 1195.81  & -65.92 &...\\ \hline
\multicolumn{14}{c}{Models for test}\\ \hline
f11E15d3 &        18.18 & 11.30 & 1.79 & 0.76 & 5.82 & 26.65  & 88.13  & 613.66  & 58.06  & 230.69  & 2202.68  & 117.88 &M7.2\\ 
f11E20d4 &        18.00 & 13.35 & 2.13 & 0.91 & 4.84 & 25.43  & 99.28  & 633.49  & 53.25  & 290.35  & 2390.70  & 109.35 &M3.2\\ 
f11E25d5 &        15.64 & 14.11 & 2.56 & 1.01 & 5.49 & 28.02  & 58.06  & 688.50  & 48.53  & 107.21  & 2357.03  & 73.19 &C1.6\\ \hline
\end{tabular}}
\begin{tablenotes}
\footnotesize
\item $^1$ Flare class below A1.0 is not labelled.
\end{tablenotes}
\end{table*}
    
   \begin{figure*}
   \centering
   \includegraphics[width=\textwidth]{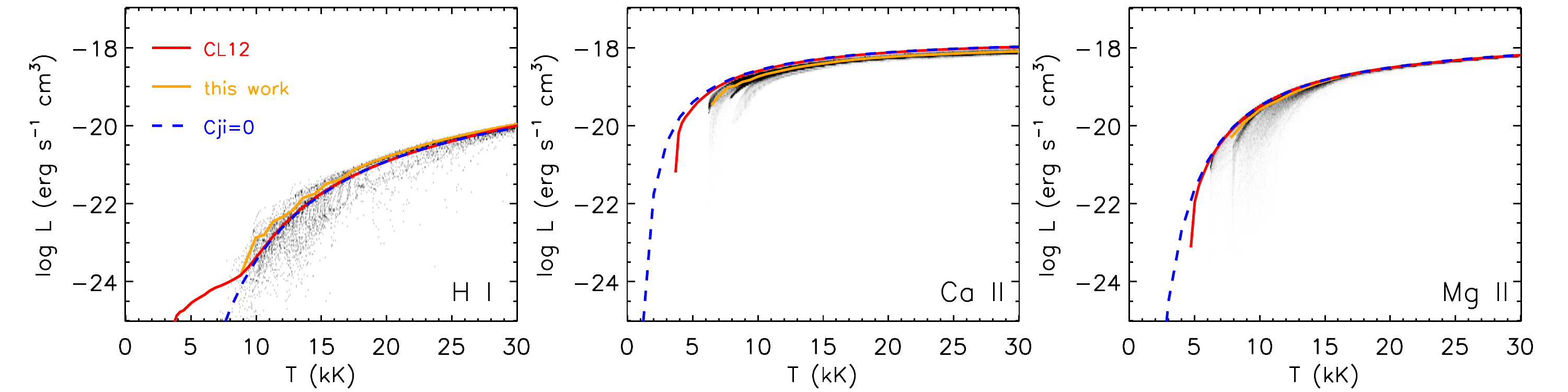}
   \caption{Probability density function of the thin radiative loss function (\ion{H}{I}, \ion{Ca}{II}, and \ion{Mg}{II}) as a function of temperature. Colored lines show relations from the recipe of \citetalias{2012carlsson} (red), the adopted fit of the PDF (yellow),   {and cases with negligible collisional deexcitation rates (blue, Eq.~(4) of \citetalias{2012carlsson}).}}
    \label{loss}
    \end{figure*}
    
   \begin{figure*}
   \centering
   \includegraphics[width=\textwidth]{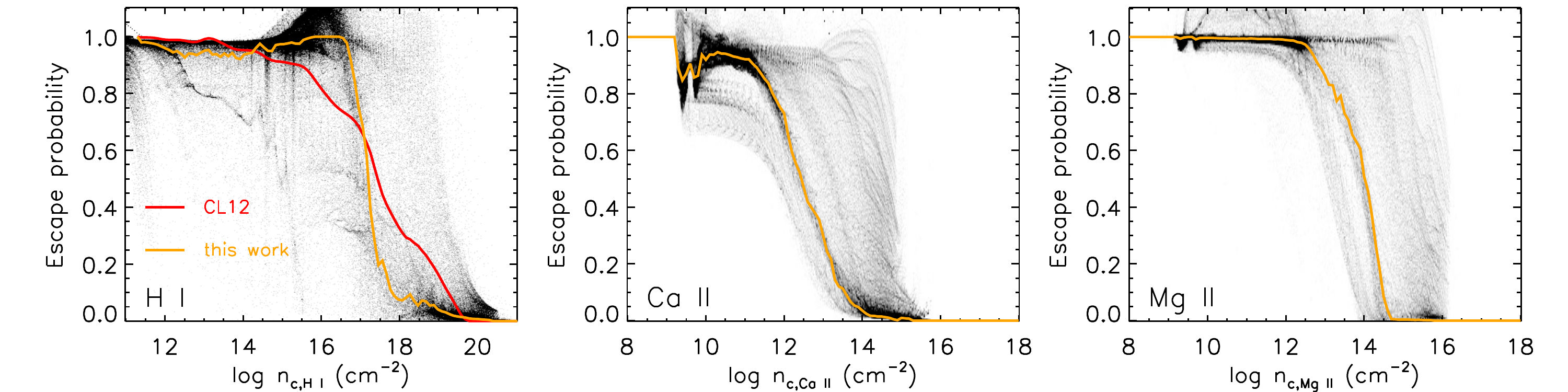}
   \caption{Probability density function of the escape probability (\ion{H}{I}, \ion{Ca}{II}, and \ion{Mg}{II}) as a function of  {the column density of the specific ion}. Colored lines show relations from the recipe of \citetalias{2012carlsson} (red) and the adopted fit of the PDF (yellow).  {Note that in the middle and right panels, the results are plotted as a function of column density of \ion{Ca}{II} and \ion{Mg}{II}, respectively, and thus we do not plot the results from the recipe of \citetalias{2012carlsson}.}}
    \label{esc}
    \end{figure*}

   \begin{figure*}
   \centering
   \includegraphics[width=\textwidth]{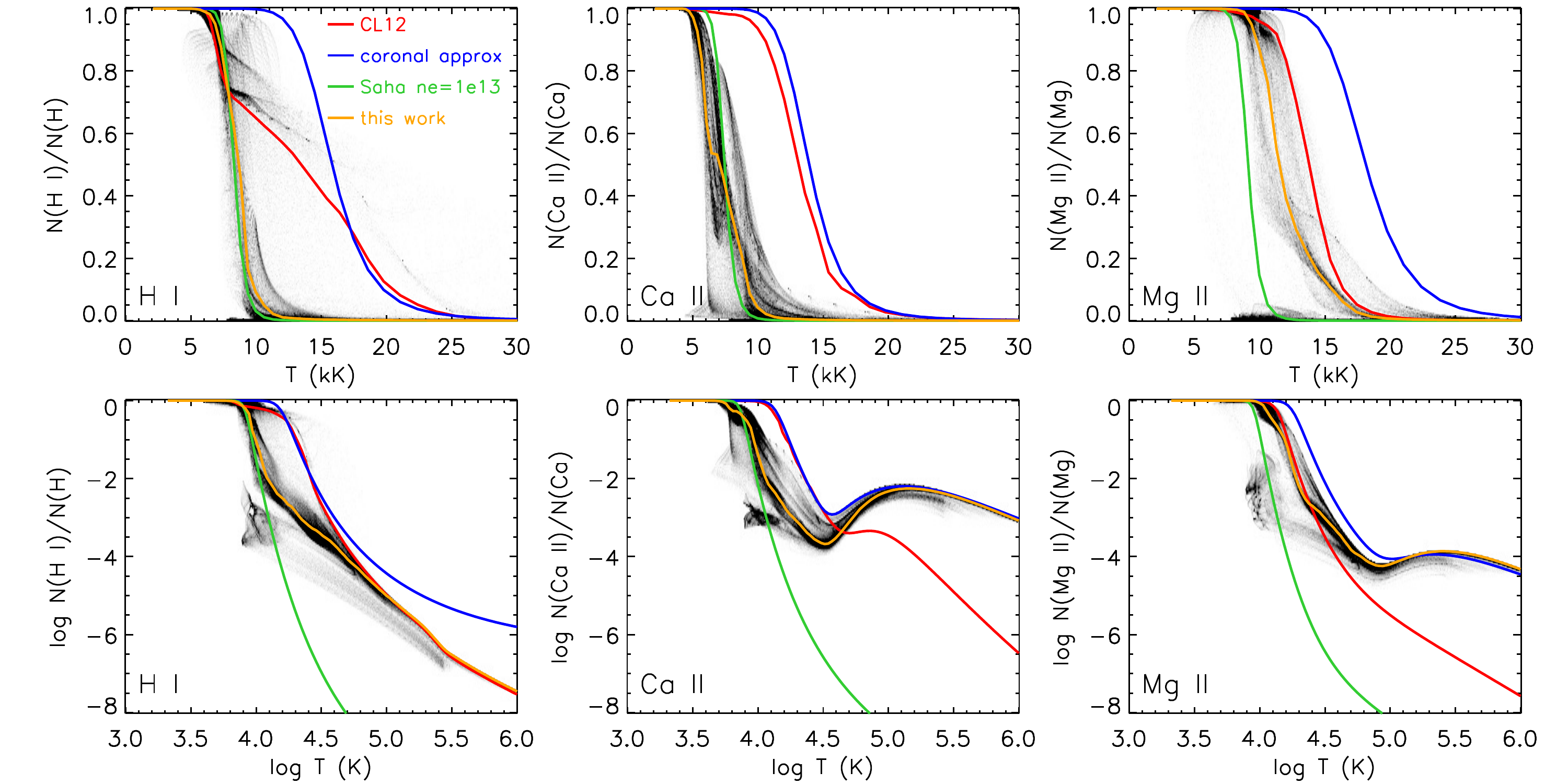}
   \caption{Probability density function of the population fractions (\ion{H}{I}, \ion{Ca}{II}, and \ion{Mg}{II}) as a function of temperature. The panels in the upper row are zoomed part of the panels in the lower row  for the temperature range of $T<30$ kK. Colored lines show the results from the recipe of \citetalias{2012carlsson} (red), the coronal approximation with a two-level atom (blue), the Saha equation with a constant $n_\mathrm{e}=10^{13}$ cm$^{-3}$ (green), and the adopted fit of the PDF (yellow).}
    \label{frac}
    \end{figure*}    
        
The recipe of \citetalias{1990gan} is constructed over only two semi-empirical models, and it is not clear whether these models can cover the variation range of solar flares. Besides, the recipe takes the height variable $z$ as an input, while in actual simulations, the height and thickness of the chromosphere are not necessarily the same as in semi-empirical models. On the other hand, the recipe of \citetalias{2012carlsson} is based on the quiet-Sun atmosphere. And it remains to be evaluated how it works during flares, since the atmospheric conditions are quite different. In this paper, we follow the steps in \citetalias{2012carlsson} and redo the fits of the three parameters from a grid of flare simulations. We briefly introduce our method in Section~\ref{sect2}. The fitting results and comparisons with other recipes are shown in Section~\ref{sect3}. We give our conclusions in Section~\ref{sect4}.

\section{Method}
\label{sect2}
We employ a grid of 25 flare models generated with the radiative hydrodynamics code \verb"RADYN" \citep{1992carlsson,1995carlsson,1997carlsson,2002carlsson}. 
 {The flare loop is assumed to be symmetric, thus half of the loop is modeled as a quarter circle with a 10 Mm length. }
 In each flare model the initial quiet-Sun atmosphere is heated by a beam of non-thermal electrons injected from the loop top \citep{2015allred}. 
 {The initial temperature at the loop top is 1 MK.}
The beam of electrons is assumed to have a power-law distribution of energy, as described with three parameters: the electron flux $F$, the spectral index $\delta$, and the cutoff energy $E_c$. The electron flux $F$ is a triangular function of time, with 10 s increase and 10 s decrease, and the peak flux is $10^{10}$ erg cm$^{-2}$ s$^{-1}$ for all models. The spectral index $\delta$ varies from 3 to 7, and the cutoff energy $E_c$ varies from 5 to 25 keV. 
   {These 25 models are used to fit the three parameters in Eq.~(\ref{cl12}), and we use another three models with a larger peak electron flux ($10^{11}$ erg cm$^{-2}$ s$^{-1}$) to test the recipe (see Sec.~\ref{sect32}).}
 {These models are labeled as f$n_1$E$n_2$d$n_3$ in Table~\ref{flux}, where the numbers $n_1$, $n_2$ and $n_3$ correspond to the values of $\log F$ at peak time, the cutoff energy $E_c$ and the spectral index $\delta$, respectively.}
 Each simulation is run for 20 s, and the snapshot at every 0.1 s is saved to the output. The \verb"RADYN" outputs are then fed into \verb"RH" \citep{2001uitenbroek,2015pereira} to calculate the \ion{Ca}{II} and \ion{Mg}{II} lines, taking into account the effect of partial frequency redistribution.  {In RH we assume statistical equilibrium.}

The H  {model} atom used in our \verb"RADYN" simulations is the same as in \citetalias{2012carlsson}, but for the Lyman series a Gaussian profile is used instead of a Voigt profile to mimic the effect of partial frequency redistribution \citep{2012leenaarts}.  The \ion{Ca}{II}  {model} atom is the same as in \citetalias{2012carlsson} and the \ion{Mg}{II}  {model} atom is the same as in \citet{2013leenaarts}. 

 {The radiative losses from H is calculated from the RADYN outputs, and the losses from Ca and Mg are calculated from the RH outputs.} We consider contribution from all the H lines between the lowest five energy levels, the Lyman continuum, the \ion{Ca}{II} H/K and triplet lines, as well as the \ion{Mg}{II} h/k lines. Balmer and higher continua of H are not considered here because they do not comply with the recipe, but we discuss in Section~\ref{sect3.4} that their contribution is not negligible and should be modeled properly. 

 {The time-averaged radiation fluxes of various spectral lines from these flare models are summarized in Table~\ref{flux}. In these models, the energy in the chromosphere is mostly dissipated through Ly$\alpha$, H$\alpha$, and \ion{Mg}{II} photons \citep{1980machado}. The flare class is estimated from the synthetic GOES 1--8 \AA\ flux that is calculated following \citet{2020kerr}, with the cross section of the flare loop assumed to be $4\times10^{15}$ cm$^2$ (a diameter of 1\arcsec). Note that in our flare models we only consider heating from non-thermal electrons, and the thermal electrons with an energy below the cutoff energy $E_c$ are all neglected. The coronal emission would be lower than expected in lack of these thermal electrons that could  heat the corona efficiently \citep{2018polito}. Therefore, the calculated soft X-ray flux, which mainly originates from coronal emissions, would be fairly underestimated if $E_c$ is large enough. Special care must be taken when comparing the flare class of the models with real observations.}

   \begin{figure*}
   \centering
   \includegraphics[width=\textwidth,trim=0 0 0 40]{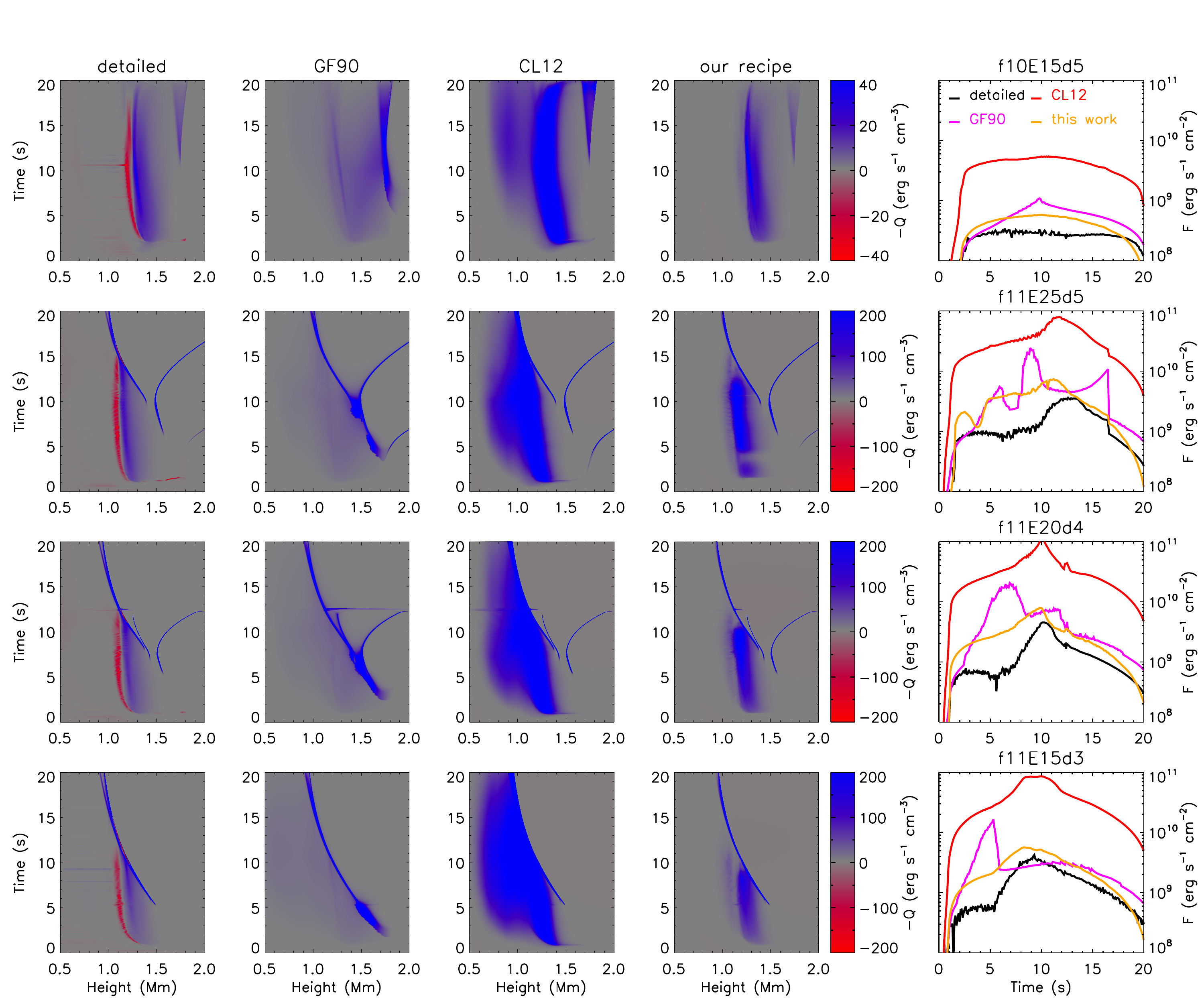}
   \caption{Comparison of chromospheric radiative losses calculated from detailed solutions and with different approximate recipes for four simulation cases. Blue colors denote radiative cooling, and red colors denote radiative heating.  {The line plots give comparisons of the evolutions of the radiative cooling rates integrated from 0.5 Mm to 2.0 Mm.}}
    \label{radloss}
    \end{figure*}

\section{Results}
\label{sect3}
\subsection{Empirical fitting results of the parameters}
As shown in Eq.~(\ref{cl12}), there are three parameters that need to be fitted empirically, the optically thin loss function $L_{X_m}$, the escape probability $E_{X_m}$, and the ionization fraction $\frac{n_{X_m}}{n_X}$.

The probability density functions (PDFs) of the optically thin loss function are shown in Fig.~\ref{loss}.  The radiative losses should be equal to the total collisional excitation from the ground state if the collisional deexcitation rates are small enough, which is the case in the quiet Sun for temperatures above 10 kK \citep{2012carlsson}. In flare cases, the fitted curves have similar shapes compared with the results from  \citetalias{2012carlsson}, despite a larger spread at low temperatures. In addition,  {the curves for cases with negligible collisional deexcitation rates (calculated from Eq.~(4) of \citetalias{2012carlsson})} and fitted curves overlap above a temperature much higher than 10 kK.

The PDFs of the escape probability are shown in Fig.~\ref{esc}. In \citetalias{2012carlsson}, the escape probability of \ion{H}{I} is tabulated as a function of the approximated optical depth of the Ly$\alpha$ line center ($\tau=4\times10^{-14}n_{c,\ion{H}{I}}$, with $n_{c,\ion{H}{I}}$ the column density of \ion{H}{I}), while the escape probability of \ion{Ca}{II} and \ion{Mg}{II} is tabulated as a function of column mass $m_c$. In flare conditions, the \ion{Ca}{II} and \ion{Mg}{II} atoms in the chromosphere are more likely to get ionized, and thus it is not appropriate to use the column mass to approximate the optical depths of these lines. Therefore, in our recipe we use the column density of \ion{H}{I}, \ion{Ca}{II} and \ion{Mg}{II} to tabulate their escape probability. There is a larger spread in the PDFs, and the curve of \ion{H}{I} escape probability is more steep than that of \citetalias{2012carlsson}. The fitted curves show small dips at low column densities, which is a result of chromospheric condensation regions where the local density is large enough to block photons to some extent.

The PDFs of the ionization fraction are shown in Fig.~\ref{frac}. It is clear that the empirical relations of \citetalias{2012carlsson} do not hold for typical  chromospheric temperatures ($T =10^4$--$10^5$ K) any more. In the flaring chromosphere, the increased electron density (in the order of $10^{13}$ cm$^{-3}$) has greatly enhanced the collisional rates, and the local atmosphere is driven towards local thermodynamical equilibrium. The fitted curves of \ion{H}{I} and \ion{Ca}{II} at low temperatures are very close to the Saha equilibrium. At high temperatures ($T>10^5$ K), the fitted curve of \ion{H}{I} follows that of \citetalias{2012carlsson}, while the fitted curves of \ion{Ca}{II} and \ion{Mg}{II} follow that of the coronal approximation with a two-level atom. The humps in the curves of \ion{Ca}{II} and \ion{Mg}{II} near $T =10^{5.2}$ K result from dielectronic recombinations.
 
   \begin{figure*}
   \centering
   \includegraphics[width=0.65\textwidth, trim=0 0 0 40]{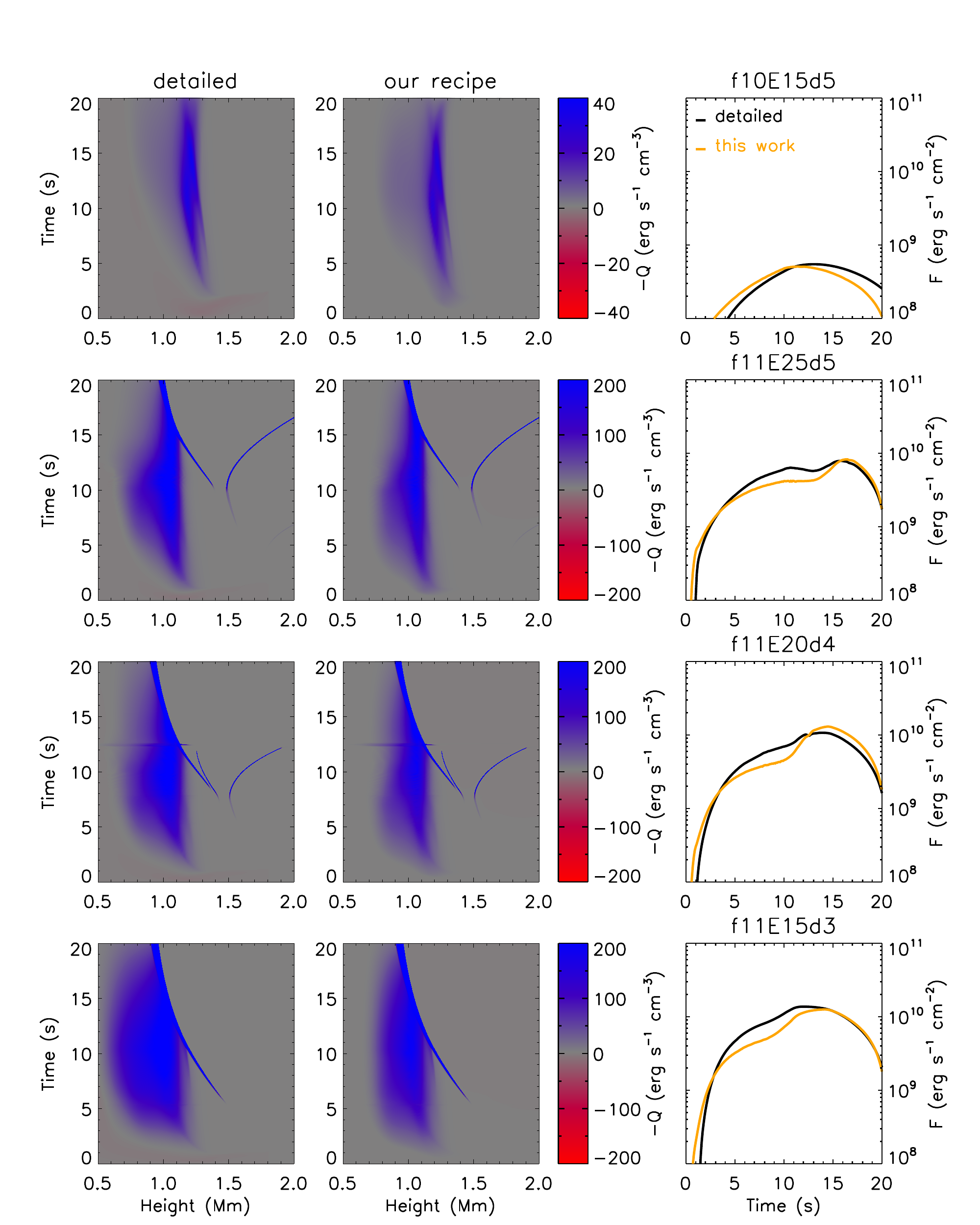}
      \caption{Same as Fig.~\ref{radloss}, but for chromospheric radiative losses contributed by \ion{H}{I} Balmer to Pfundt continua.
              }
         \label{cont}
   \end{figure*}
   
\subsection{Comparison with other recipes}
\label{sect32}
 {We choose four flare models to test the performance of different recipes. Among the four flare models, the first one (f10E15d5) is within our dataset for fitting. The peak electron flux of the other three models is one order of magnitude larger (f11, $10^{11}$ erg cm$^{-2}$ s$^{-1}$), with varying  values of $E_c$  and $\delta$ (Table~\ref{flux}).}
In Fig.~\ref{radloss} we calculate the detailed radiative losses in four flare models and compare them with the approximate results from different recipes.  { We also integrate the total radiative losses spatially (from 0.5 Mm to 2.0 Mm) and show their time evolutions.} A striking feature of Fig.~\ref{radloss} is that in spite of radiative cooling, there exists strong radiative heating in the atmosphere, which will be discussed further in Sect.~\ref{sect3.3}.

 Generally speaking, the recipe of \citetalias{1990gan} underestimates the radiative cooling in the mid-chromosphere (1.0--1.5 Mm), while in the upper chromosphere the radiative cooling is overestimated.  {The time evolution of the spatially integrated radiative cooling show multi-peaks, owing to an inaccurate estimation of the radiative cooling near the transition region.} As for the recipe of \citetalias{2012carlsson}, the radiative cooling is overestimated up to 1--2 orders of magnitude due to an overestimation of the \ion{H}{I} population fraction in the chromosphere. The results from our recipe look more reasonable, although the estimated cooling can be larger than the actual cooling by a factor of 3--5 in the f11 models (peak electron flux of $10^{11}$ erg cm$^{-2}$ s$^{-1}$).

      \begin{figure}
   \centering
   \includegraphics[width=0.35\textwidth]{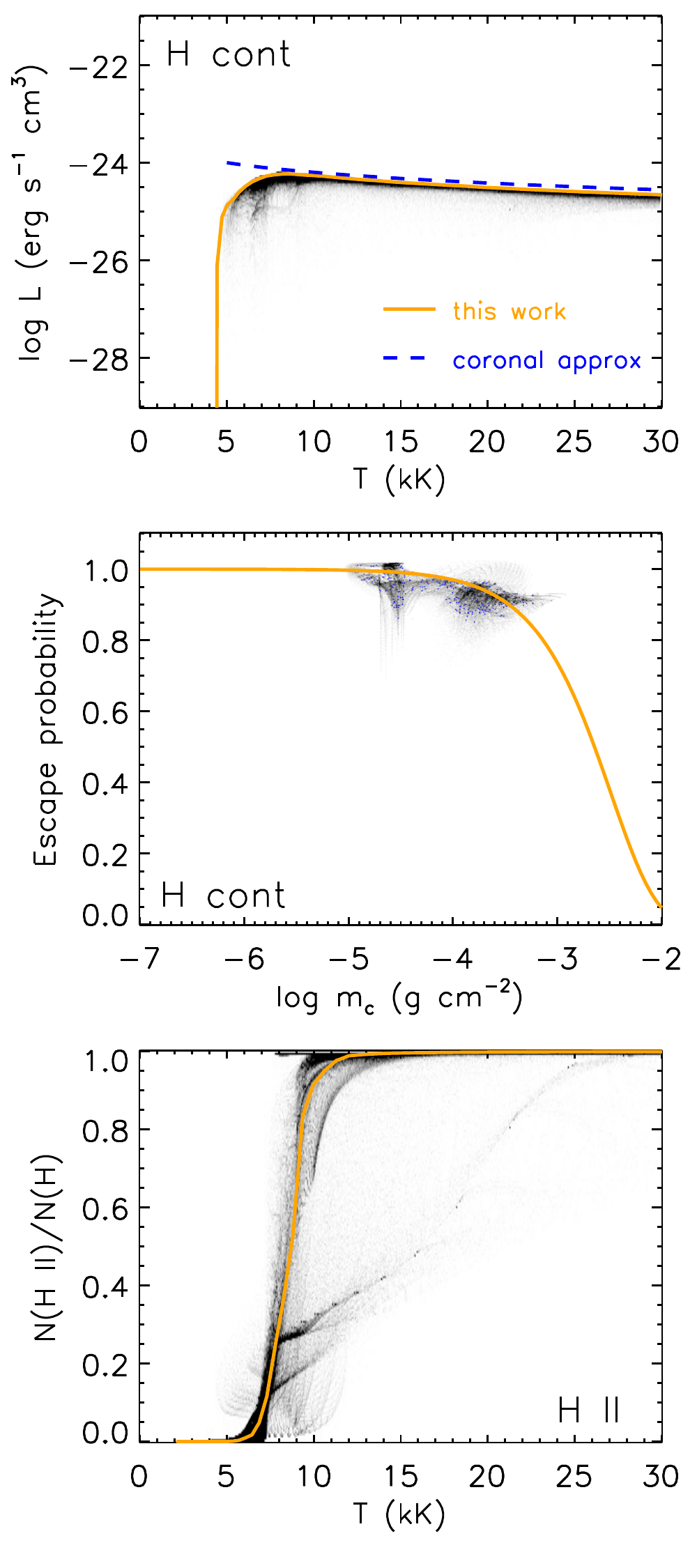}
      \caption{Probability density functions of the thin radiative loss function and the escape probability for H continua (Balmer to Pfundt), as well as the population fractions of \ion{H}{II}, as functions of temperature and column mass. Orange lines show the adopted fit,  {and the blue dashed curve in the top panel shows the thin radiative loss function under coronal approximation.}
              }
         \label{hion}
   \end{figure}
   
  {The temporally and spatially integrated total radiative losses from different recipes, as well as corresponding errors in each flare model, are also listed in Table~\ref{flux}. For most flare models, the results from our recipe are the closest to the detailed solutions. The recipe of \citetalias{1990gan} can produce an overall good approximation of the integrated total radiative losses, but the height distribution of the radiative losses is quite different from the actual one (Fig.~\ref{radloss}). In flare models with a large cut-off energy ($E_c=25$ keV) and a large spectral index ($\delta\ge5$), the radiative cooling is underestimated from our recipe, but still in the same order of magnitude compared with the exact value. This is because in these models, there would appear a specific region trapped between the transition region and the heated chromosphere, with a low temperature, a large density, and a large ionization degree, referred to as ``chromospheric bubbles'' \citep{2020reid}. According to our recipe, a low temperature would mean a low ionization degree, thus the column density of \ion{H}{I}, \ion{Ca}{II}, and \ion{Mg}{II} would be overestimated and the radiative losses are underestimated. }

\begin{table}
\caption{Total radiative fluxes of the chromosphere from H continua (Balmer to Pfundt) in different flare models.}             
\label{contflux}      
\centering
\begin{tabular}{c |  rr | r}
\hline
flare &     \multicolumn{2}{c|}{$\mathcal{F}^\mathrm{tot}_\mathrm{cont}$ ($10^9$ erg cm$^{-2}$)} & $\Delta \mathcal{F}^\mathrm{tot}_\mathrm{cont}$ (\%) \\ 
model &     detailed &  this work  & this work \\ \hline
\multicolumn{4}{c}{Models for fitting}\\ \hline
f10E05d3 &        9.00 & 8.36 & -7.09   \\ 
f10E10d3 &       11.05 & 8.55 & -22.62  \\ 
f10E15d3 &       14.59 & 10.82 & -25.88  \\ 
f10E20d3 &       17.81 & 12.91 & -27.55  \\ 
f10E25d3 &       21.48 & 15.37 & -28.44  \\ 
f10E05d4 &        5.40 & 6.33 & 17.27   \\ 
f10E10d4 &        7.42 & 7.03 & -5.22   \\ 
f10E15d4 &        8.97 & 7.76 & -13.45   \\ 
f10E20d4 &       12.24 & 10.08 & -17.66  \\ 
f10E25d4 &       16.17 & 12.92 & -20.07  \\ 
f10E05d5 &        3.76 & 4.81 & 27.80   \\ 
f10E10d5 &        6.04 & 6.44 & 6.48   \\ 
f10E15d5 &        6.08 & 5.84 & -3.88 \\ 
f10E20d5 &        8.98 & 7.93 & -11.70   \\ 
f10E25d5 &       12.56 & 10.62 & -15.57  \\ 
f10E05d6 &        2.93 & 3.93 & 34.19   \\ 
f10E10d6 &        5.40 & 6.09 & 12.83   \\ 
f10E15d6 &        4.53 & 4.60 & 1.62  \\ 
f10E20d6 &        7.10 & 6.62 & -6.87   \\ 
f10E25d6 &       10.44 & 9.11 & -12.82  \\ 
f10E05d7 &        2.48 & 3.45 & 39.01   \\ 
f10E10d7 &        5.01 & 5.83 & 16.38  \\ 
f10E15d7 &        3.78 & 3.85 & 2.11 \\ 
f10E20d7 &        5.98 & 5.77 & -3.48   \\ 
f10E25d7 &        9.10 & 8.11 & -10.84  \\ \hline
\multicolumn{4}{c}{Models for test}\\ \hline
f11E15d3 & 147.04 & 122.94 & -16.39\\
f11E20d4 & 109.77 & 110.19 & 0.38\\
f11E25d5 & 87.93 & 76.61 & -12.87\\ \hline
\end{tabular}
\end{table}

\subsection{Radiative heating}
\label{sect3.3}
As can be seen in Fig.~\ref{radloss}, there are regions where the radiative flux contributes to the internal energy as radiative heating, through absorption of photons emitted from nearby regions. These regions are always adjacent to the regions with large radiative cooling \citep{1980machado,2012carlsson}. The magnitude of radiative heating in specific regions is proportional to the magnitude of radiative cooling in the regions adjacent to them. In flares,  radiative heating is mainly contributed by Ly$\alpha$ and the Lyman continuum, as well as the resonance lines of \ion{Ca}{II} and \ion{Mg}{II}. The contribution of Ly$\alpha$ dominates at the beginning of flare heating, while at a later time, the contribution of the Lyman continuum dominates. Heating from the \ion{Ca}{II} H/K lines are partly compensated by cooling from the \ion{Ca}{II} triplet lines.

Our recipe, just as the previous ones, does not consider the contribution of radiative heating. It is a great challenge to include both radiative cooling and radiative heating together in one unified empirical formula, because it is very difficult to locate the region of radiative heating. There seems to be no quick method to determine whether the atmosphere at each height is radiatively heated or cooled. However, we do find that for flare models with the same energy cutoff $E_c$, the location of the radiative heating region is roughly the same. Therefore, for a specific $E_c$, one might get an empirical relation by fitting both positive and negative escape probabilities together, but this relation is not valid for other values of $E_c$, and it is not applicable to actual self-consistent MHD simulations. 

 {Incident radiation of the optically thin radiative losses from the corona can also heat the chromosphere \citep{1994wahlstrom,2015allred}. In flares, the magnitude of incident radiation energy could be  comparable to the amount of radiative losses from spectral lines \citep{1994hawley,2015allred}. An approximation method for this heating is described in \citetalias{2012carlsson}.}

\subsection{Cooling from Balmer and higher continua}
\label{sect3.4}
Both the above recipe and \citetalias{2012carlsson} do not include cooling from Balmer and higher continua, since they are optically much thinner than the \ion{H}{I} lines and Lyman continuum and the treatment of a unified escape probability curve is not reasonable. Their contributions to the radiative losses are shown in Fig.~\ref{cont}. For flare cases, the region with large cooling from Balmer and higher continua lies below the region with large cooling from \ion{H}{I} lines and Lyman continuum, as shown in previous calculations \citep{1986avrett,1994hawley,2019prochazka}. In these cases radiative heating from Ly$\alpha$ and Lyman continuum is compensated by radiative cooling from Balmer and higher continua to some extent. 

 {We list the values of spatially and temporally integrated radiative losses from Balmer and higher continua for different flare models in Table~\ref{contflux}.} We find that the total cooling from Balmer and higher continua can be larger than the total cooling from \ion{H}{I} lines and Lyman continuum, especially when the chromosphere is heated after a certain time \citep{1986avrett,1994hawley,2019prochazka}.  Although the recipe of \citetalias{1990gan} is constructed over the semi-empirical models, it seems that cooling from \ion{H}{I} continua is not estimated accurately. Thus it would be very important to correctly include cooling from Balmer and higher continua.

 {Radiative cooling from these continua can be approximated with Eq.~(\ref{cl12}) in a similar way. The PDFs of the thin radiative loss function and the escape probability is shown in Fig.~\ref{hion} with fitted curves. The ionization fraction of \ion{H}{II} is also fitted accordingly. The escape probability for continuum photons is close to 1 in the chromosphere, while deep down the photosphere the probability decreases to $\mathrm{e}^{-\tau}$, where $\tau$ is an approximated optical depth at these continua. We set $\tau$ proportional to column mass $m_c$: $\tau=\alpha m_c$, and a least-square best fit to the PDF gives $\alpha=3.05\times10^2$.  The calculated radiative cooling from Balmer and higher continua is also shown in Fig.~\ref{cont} and Table~\ref{contflux}, and it turns out to be a good approximation.}

\subsection{Cooling from other sources}
 {In our recipe, we only consider H, Ca, and Mg as cooling sources in the chromospheric energy balance. \cite{1989anderson} showed that in the quiet Sun, the abundant iron lines can also contribute significantly to the total radiative losses. Their contributions in flare conditions might also be important,  since the NUV \ion{Fe}{II} lines are also enhanced during flares \citep{2017kowalski,2020graham}. Other candidates include the strong \ion{He}{I} 10830 \AA\ line in flares, formed in the mid-upper chromosphere and modulated by the incident coronal radiation \citep{2005ding,2014golding,2021kerr}. Further investigations are required to quantify the contributions from these lines.}

\section{Conclusion}
\label{sect4}
In this paper, we provide a new recipe to calculate the chromospheric radiative losses based on the recipe of \citetalias{2012carlsson} but for a flaring atmosphere. We redo the fittings from a grid of flare models and tabulate the three parameters: optically thin radiative loss, escape probability, and  ionization fraction as functions of temperature or column density. The largest difference between our recipe and \citetalias{2012carlsson} lies in the empirical curve of ionization fraction. In the $10^4$ K temperature plateau of the flaring chromosphere, hydrogen is mostly ionized as a result of increased collisional ionization rates. 

The calculated radiative cooling from our recipe is a good approximation of the actual cooling in flares, especially in regions with large cooling values, while \citetalias{1990gan} tends to underestimate and \citetalias{2012carlsson} tends to overestimate the cooling rate.  {It is  noted that the \ion{H}{I} Balmer and higher continua could also  contribute significantly to the radiative cooling in flares, and they can be approximated  in a similar way. Our recipe is valid for flares with non-thermal electron peak fluxes in the range of 10$^{10}$--10$^{11}$ erg cm$^{-2}$ s$^{-1}$.}  Nevertheless, our recipe is not aimed for all kinds of solar activities, so it might not work well under atmospheric conditions that are far from flare conditions. 
  
Currently in our recipe we are unable to consider radiative heating from Ly$\alpha$ and the Lyman continuum, and it requires further study to determine how much influence there would be if radiative heating in the flaring chromosphere is neglected.

\begin{acknowledgements}
This work was supported by National Key R\&D Program of China under grant 2021YFA1600504 and by NSFC under grants 11903020, 11733003, and 12127901,  and by the Research Council of Norway through its Centres of Excellence scheme, project number 262622, and through grants of computing time from the Programme for Supercomputing.

\end{acknowledgements}

%
   \bibliographystyle{aa} 
   \bibliography{aa.bib} 

\begin{thebibliography}{32}
\expandafter\ifx\csname natexlab\endcsname\relax\def\natexlab#1{#1}\fi

\bibitem[{{Allred} {et~al.}(2015){Allred}, {Kowalski}, \&
  {Carlsson}}]{2015allred}
{Allred}, J.~C., {Kowalski}, A.~F., \& {Carlsson}, M. 2015, \apj, 809, 104

\bibitem[{{Anderson} \& {Athay}(1989)}]{1989anderson}
{Anderson}, L.~S. \& {Athay}, R.~G. 1989, \apj, 346, 1010

\bibitem[{{Avrett} {et~al.}(1986){Avrett}, {Machado}, \& {Kurucz}}]{1986avrett}
{Avrett}, E.~H., {Machado}, M.~E., \& {Kurucz}, R.~L. 1986, in The Lower
  Atmosphere of Solar Flares, ed. D.~F. {Neidig} \& M.~E. {Machado}, 216--281

\bibitem[{{Bradshaw} \& {Cargill}(2013)}]{2013bradshaw}
{Bradshaw}, S.~J. \& {Cargill}, P.~J. 2013, \apj, 770, 12

\bibitem[{{Carlsson} \& {Leenaarts}(2012)}]{2012carlsson}
{Carlsson}, M. \& {Leenaarts}, J. 2012, \aap, 539, A39

\bibitem[{{Carlsson} \& {Stein}(1992)}]{1992carlsson}
{Carlsson}, M. \& {Stein}, R.~F. 1992, \apjl, 397, L59

\bibitem[{{Carlsson} \& {Stein}(1995)}]{1995carlsson}
{Carlsson}, M. \& {Stein}, R.~F. 1995, \apjl, 440, L29

\bibitem[{{Carlsson} \& {Stein}(1997)}]{1997carlsson}
{Carlsson}, M. \& {Stein}, R.~F. 1997, \apj, 481, 500

\bibitem[{{Carlsson} \& {Stein}(2002)}]{2002carlsson}
{Carlsson}, M. \& {Stein}, R.~F. 2002, \apj, 572, 626

\bibitem[{{Del Zanna} {et~al.}(2021){Del Zanna}, {Dere}, {Young}, \&
  {Landi}}]{2021delzanna}
{Del Zanna}, G., {Dere}, K.~P., {Young}, P.~R., \& {Landi}, E. 2021, \apj, 909,
  38

\bibitem[{{Dere} {et~al.}(1997){Dere}, {Landi}, {Mason}, {Monsignori Fossi}, \&
  {Young}}]{1997dere}
{Dere}, K.~P., {Landi}, E., {Mason}, H.~E., {Monsignori Fossi}, B.~C., \&
  {Young}, P.~R. 1997, \aaps, 125, 149

\bibitem[{{Ding} {et~al.}(2005){Ding}, {Li}, \& {Fang}}]{2005ding}
{Ding}, M.~D., {Li}, H., \& {Fang}, C. 2005, \aap, 432, 699

\bibitem[{{Fang} {et~al.}(1993){Fang}, {Henoux}, \& {Gan}}]{1993fang}
{Fang}, C., {Henoux}, J.~C., \& {Gan}, W.~Q. 1993, \aap, 274, 917

\bibitem[{{Gan} \& {Fang}(1990)}]{1990gan}
{Gan}, W.~Q. \& {Fang}, C. 1990, \apj, 358, 328

\bibitem[{{Golding} {et~al.}(2014){Golding}, {Carlsson}, \&
  {Leenaarts}}]{2014golding}
{Golding}, T.~P., {Carlsson}, M., \& {Leenaarts}, J. 2014, \apj, 784, 30

\bibitem[{{Graham} {et~al.}(2020){Graham}, {Cauzzi}, {Zangrilli}, {Kowalski},
  {Sim{\~o}es}, \& {Allred}}]{2020graham}
{Graham}, D.~R., {Cauzzi}, G., {Zangrilli}, L., {et~al.} 2020, \apj, 895, 6

\bibitem[{{Gudiksen} {et~al.}(2011){Gudiksen}, {Carlsson}, {Hansteen}, {Hayek},
  {Leenaarts}, \& {Mart{\'\i}nez-Sykora}}]{2011gudiksen}
{Gudiksen}, B.~V., {Carlsson}, M., {Hansteen}, V.~H., {et~al.} 2011, \aap, 531,
  A154

\bibitem[{{Hawley} \& {Fisher}(1994)}]{1994hawley}
{Hawley}, S.~L. \& {Fisher}, G.~H. 1994, \apj, 426, 387

\bibitem[{{Kerr} {et~al.}(2020){Kerr}, {Allred}, \& {Polito}}]{2020kerr}
{Kerr}, G.~S., {Allred}, J.~C., \& {Polito}, V. 2020, \apj, 900, 18

\bibitem[{{Kerr} {et~al.}(2021){Kerr}, {Xu}, {Allred}, {Polito}, {Sadykov},
  {Huang}, \& {Wang}}]{2021kerr}
{Kerr}, G.~S., {Xu}, Y., {Allred}, J.~C., {et~al.} 2021, \apj, 912, 153

\bibitem[{{Kowalski} {et~al.}(2017){Kowalski}, {Allred}, {Daw}, {Cauzzi}, \&
  {Carlsson}}]{2017kowalski}
{Kowalski}, A.~F., {Allred}, J.~C., {Daw}, A., {Cauzzi}, G., \& {Carlsson}, M.
  2017, \apj, 836, 12

\bibitem[{{Leenaarts} {et~al.}(2012){Leenaarts}, {Carlsson}, \& {Rouppe van der
  Voort}}]{2012leenaarts}
{Leenaarts}, J., {Carlsson}, M., \& {Rouppe van der Voort}, L. 2012, \apj, 749,
  136

\bibitem[{{Leenaarts} {et~al.}(2013){Leenaarts}, {Pereira}, {Carlsson},
  {Uitenbroek}, \& {De Pontieu}}]{2013leenaarts}
{Leenaarts}, J., {Pereira}, T.~M.~D., {Carlsson}, M., {Uitenbroek}, H., \& {De
  Pontieu}, B. 2013, \apj, 772, 89

\bibitem[{{Machado} {et~al.}(1980){Machado}, {Avrett}, {Vernazza}, \&
  {Noyes}}]{1980machado}
{Machado}, M.~E., {Avrett}, E.~H., {Vernazza}, J.~E., \& {Noyes}, R.~W. 1980,
  \apj, 242, 336

\bibitem[{{Nagai}(1980)}]{1980nagai}
{Nagai}, F. 1980, \solphys, 68, 351

\bibitem[{{Pereira} \& {Uitenbroek}(2015)}]{2015pereira}
{Pereira}, T. M.~D. \& {Uitenbroek}, H. 2015, \aap, 574, A3

\bibitem[{{Polito} {et~al.}(2018){Polito}, {Testa}, {Allred}, {De Pontieu},
  {Carlsson}, {Pereira}, {Go{\v{s}}i{\'c}}, \& {Reale}}]{2018polito}
{Polito}, V., {Testa}, P., {Allred}, J., {et~al.} 2018, \apj, 856, 178

\bibitem[{{Proch{\'a}zka} {et~al.}(2019){Proch{\'a}zka}, {Reid}, \&
  {Mathioudakis}}]{2019prochazka}
{Proch{\'a}zka}, O., {Reid}, A., \& {Mathioudakis}, M. 2019, \apj, 882, 97

\bibitem[{{Reid} {et~al.}(2020){Reid}, {Zhigulin}, {Carlsson}, \&
  {Mathioudakis}}]{2020reid}
{Reid}, A., {Zhigulin}, B., {Carlsson}, M., \& {Mathioudakis}, M. 2020, \apjl,
  894, L21

\bibitem[{{Uitenbroek}(2001)}]{2001uitenbroek}
{Uitenbroek}, H. 2001, \apj, 557, 389

\bibitem[{{Wahlstrom} \& {Carlsson}(1994)}]{1994wahlstrom}
{Wahlstrom}, C. \& {Carlsson}, M. 1994, \apj, 433, 417

\bibitem[{{Wang} \& {Yokoyama}(2020)}]{2020wang}
{Wang}, Y. \& {Yokoyama}, T. 2020, \apj, 891, 110

\end{thebibliography}
%

\end{document}